# How Transit Countries Become Refugee Destinations: Insights From Central and Eastern Europe

Ciprian Panzaru[1] | Liliana Harding[2]

[1]Department of Sociology, West University of Timisoara, Timisoara, Romania | [2]School of Economics, University of East Anglia, Norwich, Norfolk, UK

**Correspondence:** Ciprian Panzaru (ciprian.panzaru@e-uvt.ro)

## ABSTRACT

This study examines how refugees' destination preferences evolve during transit, focusing on three Central and Eastern European countries—Bulgaria, Romania, and Hungary—traditionally regarded as 'transit only' prior to the Ukraine refugee crisis. Using a mixed-methods approach, we first analyse 2252 observations from the International Organisation for Migration's Flow Monitoring Surveys to identify the main factors influencing changes in destination choices. We then complement these findings with qualitative data from focus groups with 16 asylum seekers in Romania to explore these dynamics in depth. Our results show that prioritising safety significantly increases the likelihood of asylum seekers reconsidering a transit country as a potential destination. Other influential factors include asylum conditions, migration costs, and educational background, with more educated individuals more likely to revise initial plans. Although our primary focus is on asylum seekers, we find that high migration costs also affect decisions, suggesting a need to 'recover' investments through settlement in more stable or economically attractive countries. The qualitative findings support the quantitative results, highlighting the role of legal stability, social networks, and perceived opportunity in shifting preferences. Overall, the study suggests that under certain conditions, transit countries can become viable destinations and supports the application of bounded rationality and human capital theory in understanding refugee decision-making.

## 1 | Introduction

The rapid rise in global refugee flows has placed migration at the forefront of political and public discourse, sparking debates over the legitimacy and governance of different types of movement. Much of this debate focuses on the motivations behind forced migration, particularly among those fleeing war, persecution, or instability. While push factors have been widely studied (Adhikari 2011; Bohra-Mishra and Massey 2011; Crippa et al. 2024; Seven 2022), less is known about how destination preferences evolve as refugees move through multiple countries. Migration decisions are not static; rather, they reflect a dynamic interplay of factors encountered along the route, including legal environments, reception conditions, social networks, and access to protection. As a result, the asylum seekers may revise their plans in response to changing security, economic, and legal circumstances (Akesson and Coupland 2018; Day and White 2002; Manafi and Roman 2022; Segal 2021; Zimmermann 2010). Although the importance of safety is well established, the role of transit-country conditions on final settlement decisions warrants further exploration.

Despite growing scholarly interest in refugee migration across Europe, limited research has examined the decision-making processes of asylum seekers transiting through countries traditionally regarded as waypoints rather than destinations. Central and Eastern European states, particularly Romania, Hungary, and Bulgaria, have historically occupied this role. However, recent developments have begun to challenge this view. Following the Russian invasion of Ukraine in 2022, all three countries received significant numbers of displaced persons. As of early 2025, Romania hosted nearly 180,000 Ukrainian refugees, Bulgaria over 200,000, and Hungary more than 62,000 (UNHCR, n.d.). Notably, these countries were already managing substantial migration flows during the 2015–2016 refugee crisis, primarily from Syria, Afghanistan, Iraq, and Pakistan. Hungary experienced the most dramatic surge, receiving over 177,135 asylum applications in 2015 due to its location along the Western Balkans route. Bulgaria saw applications rise sharply from 7145 in 2013 to 20,390 in 2015 (Eurostat, n.d.). Although Romania historically received fewer applications (European Commission 2021), it nonetheless experienced a noticeable increase in inflows during the same period, becoming a key transit route for asylum seekers travelling from the Middle East and Africa toward Western Europe. While these numbers were lower than those recorded in major host countries—Germany registered over 745,000 applications in 2015 alone—they still placed considerable pressure on national asylum systems, revealing both infrastructural gaps and divergent policy responses.

This study investigates the decision-making processes of asylum seekers in Romania, Hungary, and Bulgaria, countries historically perceived as transit zones but now increasingly involved in long-term refugee reception. We examine how asylum seekers revise their destination preferences in response to legal, social, and economic conditions encountered in transit. Using a mixed-methods approach, we draw on 2252 observations from the International Organisation for Migration's (IOM) Flow Monitoring Survey and qualitative data from two focus groups conducted at the Regional Centre for Procedures and Accommodation for Asylum Seekers in Timișoara, Romania. Romania was selected for the qualitative analysis to provide a focused case study of how transit-country experiences can lead to long-term settlement. The study also draws on longstanding collaborations with local organisations that have supported asylum seekers in Timișoara since the 2015–2016 refugee crisis, enabling us to test theory-driven insights from the quantitative analysis in a real-world context.

Throughout this study, we maintain a clear distinction between asylum seekers—those awaiting decisions on their protection status, and refugees who have been formally recognised. Asylum seekers often face greater legal uncertainty and more limited access to services, which may significantly influence how they navigate transit and destination contexts. Given that this study focuses on asylum seekers residing in an asylum centre, we acknowledge that their experiences may differ from those of recognised refugees and may influence how they navigate both transit and destination contexts. Nevertheless, examining how asylum seekers construct destination choices under such constrained conditions offers important insights into forced migration decision-making under uncertainty.

This paper proceeds as follows. We begin with a review of the relevant literature and theoretical frameworks. We then present findings from our logistic regression analysis using IOM survey data, followed by qualitative insights from focus group discussions in Romania. Finally, we discuss the implications of our mixed-methods analysis for understanding how transit countries may evolve into refugee destinations.

## 2 | Literature Review

Refugee migration in Europe has long been a complex and dynamic phenomenon, driven by conflict, persecution, and environmental crises. While much scholarly attention has focused on Western European states, the role of Central and Eastern European countries, including Romania, Bulgaria, and Hungary, has become increasingly significant in recent years. Historically regarded as transit zones rather than final destinations, these countries have been central to key migration routes, particularly during the 2015–2016 refugee crisis. In response to rising inflows, Hungary and Bulgaria implemented restrictive asylum measures, including border closures and deterrence practices, whereas Romania began enhancing its reception infrastructure. These divergent policy responses have shaped the region's capacity to manage refugee flows. In this evolving context, understanding how refugees make destination choices is critical. Although their backgrounds vary widely, several factors shape these decisions, including legal status, access to protection, public perceptions, economic prospects, and social networks (Crawley and Hagen-Zanker 2019; Hofmann 2015; Manafi and Roman 2022; Neumayer 2004).

Network theory offers one of the most established frameworks for explaining refugee decision-making. According to Massey et al. (1993), migrants rely on pre-existing social networks comprising family, friends, or community members who have already migrated. These networks provide critical resources such as information, financial support, and assistance with integration and reduce the risks and costs associated with migration. Social networks—whether physical or digital—also enable real-time information sharing via platforms like social media, further reinforcing their influence on destination choices (AbuJarour et al. 2021; Miconi 2020).

Human capital theory complements this view by focusing on how migrants aim to maximise their economic potential. Refugees with higher levels of education or professional experience tend to prefer destinations that offer favourable labour market conditions and greater opportunities for upward mobility (Borjas 1987; Dustmann and Weiss 2007). Conversely, those with fewer skills may target destinations with more accessible low-skilled employment, adjusting their plans based on local economic conditions.

Bounded rationality theory further deepens our understanding by recognising the limited information, uncertainty, and constraints that shape refugee decisions. Rather than always pursuing optimal outcomes, refugees often settle for options that meet minimum thresholds of safety and opportunity (Brunarska 2019; Simon 1957). For example, changing legal barriers, economic hardships, or evolving personal circumstances can prompt asylum seekers to revise their original plans and remain in transit countries. Kuschminder and Waidler (2020) underscore this through their study of refugees in Greece and Turkey, showing how transit country conditions such as employment, language, and housing can heavily influence onward migration decisions.

Alongside bounded rationality, migration risks play a significant role in shaping refugee behaviour. Refugees face cumulative risks including violence, exploitation, and legal uncertainty, and must continuously evaluate whether to move forward or remain in relatively safer environments (Becker and Ferrara 2019). The decision to stay in a transit country can thus be understood as a risk-avoidance strategy, especially when the path forward carries high levels of danger.

Economic factors also weigh heavily in refugee decisions. Aksoy and Poutvaara (2019) highlight the role of economic self-selection, whereby individuals seek destinations that offer employment prospects and economic stability. The availability of jobs, ease of obtaining work permits, and local economic conditions often determine whether refugees continue their journey or settle where they are. For more skilled refugees, stable employment in transit countries may offset the risks of moving further.

Host society attitudes also shape decision-making. Research suggests that Ukrainian refugees, for instance, have received significantly different treatment compared to asylum seekers from the Middle East or Africa, due to cultural similarity and perceived lower threat levels (Kossowska et al. 2023; Sinclair et al. 2024). Such distinctions have implications for how transit countries become viable destinations for some groups but remain temporary stopovers for others.

Institutional theory adds another dimension by highlighting the role of migration and asylum policies in shaping refugee preferences. Countries with more favourable asylum frameworks, legal protections, and welfare systems may be more attractive to refugees (Freeman 2004; Hollifield 2004). However, evidence is mixed regarding the actual impact of welfare benefits. While some studies suggest these play a secondary role to legal status or family reunification (Diop-Christensen and Lanciné 2022), others show context-dependent effects (Dellinger and Huber 2021; Ferwerda et al. 2022). Kuschminder and Koser (2017) specifically emphasise how policies in transit countries such as work rights and legal protections can strongly influence the decision to remain or move onward.

Our research builds on the theoretical insights outlined above. Based on these, we hypothesise the following: (1) safety and legal stability in transit countries increase the likelihood that refugees will treat a transit country as a final destination; (2) refugees with higher levels of human capital (e.g., education or skills) are more likely to alter their original plans when presented with favourable opportunities in transit countries; and (3) strong pre-existing social ties at the originally intended destination reduce the likelihood of switching to a transit country.

## 3 | Data and Methods

This study employed a mixed-methods approach, integrating both quantitative and qualitative data to gain a comprehensive understanding of the drivers of migration and the patterns of refugees' clustering in specific destinations or transit countries. The primary research question guiding the study sought to explore how forced migrants select their destination countries and how these preferences evolve when they pass through or settle in locations along the migration route. The quantitative analysis identifies *general drivers of changing destination preferences* among transit-country asylum seekers, while the qualitative focus groups probe *why* and *how* those drivers come into play in individuals' lived experiences.

### 3.1 | Quantitative Analysis

The quantitative data for this study were sourced from the Flow Monitoring Surveys (FMS), conducted by the International Organisation for Migration (IOM) as part of its broader efforts to monitor and analyse migration along key routes. The FMS gathers data through structured surveys and interviews with migrants at major transit points, border crossings, and reception centres, offering detailed insights into migration patterns, individual trajectories, and the evolving dynamics of population movement. For the purposes of this study, the FMS provides information on asylum seekers' individual characteristics, migration trajectories up to the point of interview, the monetary cost of migration, social ties at both destination and transit points, and, crucially, destination preferences at both the origin and transit stages. This dataset is particularly well suited for examining migration through countries which lie along key migration routes and are often categorised as transit countries.

In this study, we apply quantitative analysis to FMS data focusing on three understudied transit contexts in Central and Eastern Europe: Bulgaria, Hungary, and Romania. The data collection process for the reference countries varied slightly: surveys for Hungary and Bulgaria were conducted between February and August 2017, while for Romania, the data were collected between August and September 2017. The reference period was marked by significant political and economic changes, including the conflict in Syria and the growing concentration of displaced populations in earlier transit locations. Availability of data for different indicators differs slightly by country, and our main analysis includes variables recorded evenly across the three transit countries.

To explore the factors influencing refugees' choice of destination, we applied the forced migration framework developed by Aksoy and Poutvaara (2019). This model builds on Borjas (1987) human capital theory of migration, which emphasises how migrants seek to maximise their economic potential, and adds risks related to conflict and persecution, as highlighted by Becker and Ferrara (2019). By integrating both economic utility and migration risks, the framework offers a more nuanced understanding of how refugees' preferences evolve during their journey, aligning with the concept of bounded rationality discussed in the literature review.

To quantify migration decisions, we applied a formal model that calculates the expected utility of remaining in the origin country versus migrating to a destination and incorporating both economic considerations and migration risks. The following equations outline how this expected utility is determined:

$$\mathrm{EU}_i^d > \mathrm{EU}_i^k \tag{1}$$

The expected utility ($\mathrm{EU}_i^k$) for an individual $i$ from an origin country $k$ to remain in their home country is modelled as a function of their earnings potential in the home economy ($w_i^k$), the home country-specific risk (utility loss at home) $q_k$ of loss of income $w_i^k$ and the utility loss $L_k$ (due to conflict and war) in the country of origin $k$:

$$\mathrm{EU}_i^k = \left(1 - q_k\right) \log\left(w_i^k\right) - q_k L_k \quad (2)$$

For migration to a destination country $d$, the expected utility is modelled as follows:

$$\mathrm{EU}_i^d = \left(1 - s_k\right) \log\left(w_i^d\right) - s_k L_M - D_i c_k + \varepsilon_i \quad (3)$$

In this equation, the expected utility $\mathrm{EU}_i^d$ is influenced by the risk $s_k$ of not reaching the intended destination $d$, the wage that can be earned in the preferred destination country $w_i^d$ and additional losses $L_m$ encountered along the migration path (or becoming trapped in transit). Individual characteristics of migrants $D_i$ and differential migration costs, for example related to education level,[1] $c_k$ further influence the expected utility of migration. Ultimately, the destination choice in the model is informed by the highest expected utility of migrating to a specific destination, compared to remaining in the country of origin.

Drawing from this framework, we matched the variables available in the IOM survey with the different elements summarised by the utility framework along categories capturing gains and losses at alternative destinations and along the path of migration, as well as individual differences and migration circumstances. That allows us to construct an empirical strategy built on a logistic regression, including the factors most significantly associated with refugees' destination preferences and the likelihood of changes in preferred destinations as a dependent variable. The logistic equation took the form of: $p = 1 / \left(1 + e^{-Z}\right)$ where $p$ will be the probability of *changing the preferred destination from the beginning of migration in the country of origin to a new destination while in transit*, at the point of the IOM survey. $Z$ represents a function of all factors influencing the destination choice made by migrants and captured by the IOM survey and as detailed below.

In sum, the IOM survey provides a robust set of variables for testing the hypotheses developed from migration theory and utility-based decision models. The dataset enables us to capture both enabling factors such as safety, legal conditions, and social ties, as well as constraining factors, including financial costs and limited choice, offering a comprehensive view of how transit country contexts shape evolving refugee destination decisions.

### 3.2 | Qualitative Data

Qualitative data for this study were collected through two focus groups involving refugees residing in Romania at the time of the research. Participants for the focus groups were selected from the Regional Centre for Procedures and Accommodation for Asylum Seekers in Timișoara, a facility operated by the Romanian Immigration Office. The centre offers accommodation and support to asylum seekers who arrive in Romania and lack the financial means to live independently. Residents of the centre are typically hosted until their asylum applications are processed by Romanian authorities. During this time, neither children nor adults are formally enrolled in education programmes. Yet, a range of non-governmental organisations (NGOs) offer educational and training initiatives, such as language courses. The centre operates under an open regime, allowing residents the freedom to travel outside the facility. Asylum seekers are granted temporary identity documents, which confirm their legal right to stay in Romania and, conditionally, to work, pending the approval of their refugee status.[2]

Participants were selected using convenience sampling, based on their availability to participate in the study. Recruitment was facilitated by LOGS,[3] a community support organisation based in Timișoara, ensuring voluntary participation without any perceived influence on asylum applications.

Since conducting research with asylum seekers requires careful ethical consideration due to their legal uncertainty, dependence on state authorities, and heightened vulnerability (Stewart 2005), this study strictly adhered to established ethical guidelines. Participants' identities were anonymised to protect their privacy, and verbal consent was obtained from all respondents. This process acknowledged both the political sensitivity and the personal risks involved for individuals who have been forcibly displaced from their countries of origin. The focus group discussions were conducted in Pashto and Dari, with the assistance of an interpreter who was also an asylum seeker and had previous experience working as a translator for local organisations. The questions were posed in English and translated into the participants' native languages. To ensure accuracy and minimise potential biases in translation, a summary of the key discussion points was presented to the participants at the conclusion of each focus group, allowing them to confirm or clarify the information recorded. Interviewers were trained to minimise retraumatisation and provided participants with information on available support services. These precautions align with best practices in refugee research (Jacobsen and Landau 2003), reinforcing a 'do no harm' approach when working with vulnerable populations.

The sample reflected the diverse nationalities present in Romanian refugee reception centres, with respondents originating from various countries. Our qualitative sample comprised only male respondents, reflecting the gender composition in the asylum centre at the time of the study. Two focus groups, with a total of 16 participants, were conducted between April and May 2021. Details of participant demographics are presented in the Appendix S1.

The focus group discussion guide was structured around themes drawn from our theoretical framework. We asked participants questions aimed at how they make choices under uncertainty (to reflect bounded rationality), what role their skills and aspirations play (human capital), and how social connections influence their decisions (network theory). Key questions included: (1) 'How would you describe the political, economic, or social situation in your home country?', (2) 'What is life like in your home country, particularly in terms of freedoms?', (3) 'What are the perceptions of migration in your home country?', (4) 'When did you decide to leave your country?', (5) 'How did you leave your country?', (6) 'What was your intended destination?', (7) 'What were the most challenging aspects of your journey, and how long did it take?', (8) 'What were your experiences crossing borders in different countries?', (9) 'What has your experience in Romania been like?', and (10) 'What are your future plans?', along with discussing whether participants intended to stay in Romania or move to another destination.

## 4 | Results and Discussion

In the first stage of analysis, Flow Monitoring Survey (FMS) data for Hungary, Bulgaria, and Romania were analysed using logistic regression to identify the factors influencing refugees' destination preferences across the selected transit countries. In the next stage, the quantitative findings from Romania were supplemented with a qualitative approach, based on focus group interviews to explore the in-depth motivations of refugees' migration choices.

### 4.1 | Regression Analysis Results

A total of 2252 migrants are included in our sample, as surveyed by the International Organisation for Migration (IOM) in Romania (14.7%), Hungary (19.18%), and Bulgaria (66.12%) between April and August 2017 (IOM 2017, 2018). The majority of respondents were male (71.4%), with the largest groups originating from Afghanistan (32.06%), Pakistan (24.29%), Syria (21.31%), and Iraq (15.14%). The average age of the respondents was 28 years, and the majority were adult males (81%). Additionally, 7% of the IOM survey respondents were children aged between 14 and 17 years.

Nearly half of the survey respondents (48.2%) had completed secondary education, and the majority were single (65.1%). 29.8% of the sample reported having children. 31.31% were employed at the time of leaving their origin, while a significant portion (44.14%) were unemployed. The estimated cost of migration reported by the migrants surveyed varied between $2500 and $5000 (38.9%), with a substantial proportion (37.37%) indicating that they had spent more than $5000 on their journey.

The most common migration route reported was through Turkey, followed by Greece and Serbia. The primary reasons for migration were war or conflict (78.8%), with economic factors (16.5%) and violence or persecution (13.9%) as further drivers. The most frequently cited intended destination at the start of migrants' journey was Germany (29.9%), followed by Italy (12.1%)—with changes of preferred destination being cited along the journey. This change in preferred destination is a primary focus of our quantitative analysis, as discussed below (a series of figures derived from the descriptive statistics are provided in Appendix S4, illustrating key variables such as migration costs, education levels, change in destination).

The IOM data made it possible to compare migrants' stated destination preferences at two key moments: (1) upon departure from their country of origin, and (2) at the time of the interview conducted in one of the three transit countries—Romania, Bulgaria, or Hungary. Based on this comparison, we constructed a binary dependent variable indicating whether the respondent had changed their preferred destination during the migration journey. This variable forms the outcome in a series of logistic regression models aimed at identifying the factors associated with such preference changes upon arrival in a transit country. Two types of changes in preferences were examined: (1) a change in preference to migrate to any destination other than the originally preferred one, and (2) a change in preference to settle in the current host (or 'transit') country—distinct from the initial destination preference. The model includes the following predictor variables: whether the respondent fled war (war refugee), perceived asylum opportunities at the originally intended destination, presence of close relatives or co-nationals at that destination, prioritisation of safety over economic conditions, education level (high vs. low), employment status in the country of origin, whether the respondent self-financed their journey, and whether the migration cost exceeded $5000. Additionally, variables capturing respondents' motivations such as reporting 'only choice' or 'other' as the basis for their destination choice were included to reflect constrained or less structured decision-making.

The results of these models are presented in Table 1. Column (1) includes the results of a pooled logit specification, and Column 2 adds transit (or survey) country fixed effects.

The regression analysis reveals several key factors associated with changes in refugees' destination preferences. Notably, originating from a country affected by war is significantly and

**TABLE 1** | Switching destination preferences—pooled and conditional (transit country fixed effects) logit regression results.

| Variable | Pooled logit B (SE) | | Conditional logit B (SE) | |
|---|---|---|---|---|
| War at origin | −0.41*** | (0.11) | −0.47*** | (0.12) |
| *Asylum* | *0.10* | *(0.12)* | *0.22** | *(0.13)* |
| *Relatives* | *−1.83**** | *(0.19)* | *−1.65**** | *(0.20)* |
| *Co-nationals* | *0.29* | *(0.44)* | *0.43* | *(0.46)* |
| *Only choice* | *1.28**** | *(0.15)* | *0.80**** | *(0.17)* |
| *Other* | *0.94**** | *(0.21)* | *1.23**** | *(0.21)* |
| *Safety* | *−0.88**** | *(0.17)* | *−0.50*** | *(0.18)* |
| Education (high vs. low) | 0.23** | (0.11) | 0.27** | (0.11) |
| Employed at home | 0.14 | (0.10) | 0.15 | (0.10) |
| Self-financed | −0.10 | (0.10) | −0.47*** | (0.11) |
| Cost > $5000 | 0.21** | (0.10) | 0.15 | (0.10) |
| Constant | 0.15 | (0.13) | — | |
| Log likelihood | −1378.2 | | −1302.95 | |
| Pseudo $R^2$ | 0.117 | | 0.089 | |
| Observations (*N*) | 2252 | | 2252 | |

*Note:* Values are coefficients with robust standard errors in parentheses. The results in 'italic' compare the motivation for a particular destination with the socio-economic conditions. *$p < 0.10$, **$p < 0.05$, ***$p < 0.01$.

negatively associated with the likelihood of switching destination preferences ($\beta = -0.41$, SE = 0.11, $p < 0.01$, pooled; $\beta = -0.47$, SE = 0.12, $p < 0.01$, fixed effects). In contrast, asylum conditions and the presence of unrelated co-nationals at the intended destination do not appear to significantly influence these preferences ($\beta = 0.10$, SE = 0.12, $p = 0.41$, pooled; $\beta = 0.22$, SE = 0.13, $p < 0.10$, fixed effects). The presence of close relatives in the originally intended destination is consistently and strongly associated with a reduced likelihood of switching destinations ($\beta = -1.83$, SE = 0.19, $p < 0.01$, pooled; $\beta = -1.65$, SE = 0.20, $p < 0.01$, fixed effects).

Prioritising safety over economic opportunity is also significantly and negatively associated with preference changes ($\beta = -0.88$, SE = 0.17, $p < 0.01$, pooled; $\beta = -0.50$, SE = 0.18, $p < 0.05$, fixed effects). Furthermore, those indicating 'no other option' as their primary motivation for choosing a destination are significantly more likely to have changed preferences, suggesting a reactive decision-making process possibly driven by structural constraints in transit contexts ($\beta = 1.28$, SE = 0.15, $p < 0.01$, pooled; $\beta = 0.80$, SE = 0.17, $p < 0.01$, fixed effects). Likewise, citing 'other' reasons for destination choice also shows a consistent and positive association with switching behaviour ($\beta = 0.94$, SE = 0.21, $p < 0.01$, pooled; $\beta = 1.23$, SE = 0.21, $p < 0.01$, fixed effects).

Importantly, higher levels of education, categorised here as 'high' (upper secondary or above) versus 'low' (up to lower secondary) are positively associated with a greater likelihood of changing destination preferences, with statistically significant effects in both models ($\beta = 0.23$, SE = 0.11, $p < 0.05$, pooled; $\beta = 0.27$, SE = 0.11, $p < 0.05$, fixed effects). Pre-migration employment shows a positive but statistically insignificant association with preference changes ($\beta = 0.14$, SE = 0.10, $p = 0.16$, pooled; $\beta = 0.15$, SE = 0.10, $p = 0.14$, fixed effects).

Self-financed migration is negatively associated with changes in destination preferences, and this association is statistically significant in the fixed effects model only ($\beta = -0.10$, SE = 0.10, $p = 0.33$, pooled; $\beta = -0.47$, SE = 0.11, $p < 0.01$, fixed effects). Finally, higher migration costs, measured as expenditures exceeding $5000, are positively associated with switching preferences in the pooled model ($\beta = 0.21$, SE = 0.10, $p < 0.05$), but the effect is not statistically significant in the fixed effects model ($\beta = 0.15$, SE = 0.10, $p = 0.14$). This discrepancy may reflect heterogeneity across countries of transit and varying positions along the migration trajectory, which are captured by the fixed effects specification.

These findings suggest that refugees' destination preferences are shaped dynamically during transit, responding to both personal constraints and contextual conditions. Notably, motivations tied to limited choice and other non-economic considerations, such as perceived safety and situational necessity, emerge as strong positive predictors of preference change. Higher levels of education are also associated with an increased likelihood of altering destination plans. In contrast, having close relatives in the originally intended destination consistently deters preference shifts, likely reflecting stronger social anchoring. This supports network theory, as strong family ties (but not merely having compatriots around) significantly discourage changes in destination choice. Moreover, while self-financed migration does not appear to significantly influence switching behaviour in the pooled model, it exhibits a substantial negative effect under fixed effects, indicating that financial investment may reduce flexibility in decision-making once transit has begun.

In the second stage of the regression analysis, we focused exclusively on asylum seekers surveyed in Romania to assess whether their destination preferences had changed upon arrival, compared to their originally intended destinations. This country-specific focus allowed us to generate more contextualised insights and align the quantitative findings more closely with the qualitative research conducted in the same setting. The Romanian subsample included 324 male respondents, primarily from Iraq (48.03%) and Syria (38.36%). Most participants were single (50.76%), childless (59.82%), had completed secondary education (40.48%), and 40.79% reported prior employment. The same set of explanatory variables used in the pooled regression model was applied to the Romanian sample. Descriptive statistics for this subsample are summarised in Appendix S1, and the regression results are presented in Table 2. An equivalent analysis for Hungary and Bulgaria is provided in Appendix S2.

The results indicate that prioritising safety over economic conditions in a destination is a strong and statistically significant predictor of changing preferences to Romania as the intended destination ($\beta = 4.44$, SE = 0.68, $p < 0.01$). Additionally, asylum conditions ($\beta = 2.20$, SE = 0.72, $p < 0.01$) and having no alternative other than to choose Romania ('only choice') ($\beta = 3.14$, SE = 1.38, $p < 0.05$) are also significant factors positively associated with preference change. In contrast, self-financed

**TABLE 2** | Switching destination preference to Romania—logit regression results.

| Variable | *B* | SE | *p* |
|---|---|---|---|
| War at origin | −0.41 | 0.64 | 0.519 |
| *Asylum* | *2.20*** | *0.72* | *0.002* |
| *Relatives* | *−0.69* | *0.80* | *0.387* |
| *Co-nationals* | *1.23* | *0.94* | *0.191* |
| *Only choice* | *3.14** | *1.38* | *0.023* |
| *Safety* | *4.44*** | *0.68* | *0.000* |
| Education (high vs. low) | −0.08 | 0.49 | 0.871 |
| Employed at home | 0.63 | 0.46 | 0.169 |
| Self-financed | −1.16** | 0.51 | 0.022 |
| Cost > $5000 | 0.84* | 0.48 | 0.08 |
| Constant | −3.08*** | 0.84 | 0.000 |
| Log likelihood | −75.13 | | |
| Pseudo $R^2$ | 0.425 | | |
| Observations (*N*) | 324 | | |

*Note:* Coefficients (*B*) reported with robust standard errors (SE). *$p < 0.10$, **$p < 0.05$, ***$p < 0.01$.

migration is negatively associated with the likelihood of preferring Romania as a destination ($\beta = -1.16$, SE $= 0.51$, $p < 0.05$), suggesting that individuals who cover their own migration costs may be less likely to settle in Romania. Migration costs exceeding $5000 show a marginally significant and positive association with preference change ($\beta = 0.84$, SE $= 0.48$, $p < 0.10$), potentially reflecting the sunk costs of prolonged journeys and the economic rationality of settling rather than continuing migration.

Based on the significant negative effect of self-financing on the intention to stay in Romania, we argue that having incurred high direct individual costs acts as a deterrent to remaining in Romania. This finding suggests that self-financed migrants may prefer to continue their journey toward higher-income destinations. In contrast to patterns observed in the pooled sample, education plays a minimal role in explaining destination preference changes in the Romanian subsample. In this specification, the binary indicator for higher education is not statistically significant ($\beta = -0.08$, SE $= 0.49$, $p = 0.87$). Alternative estimations using multiple education levels yielded consistent results, reinforcing the conclusion that, in the Romanian context, educational attainment does not substantially influence the likelihood of switching preferences.

### 4.2 | Results From Qualitative Analysis

The qualitative component of this study draws on two focus groups conducted in April–May 2021, with a total of 16 participants. The participants' mean age was 24.12 years (standard deviation $= 6.48$), and most originated from Pakistan ($N = 10$), followed by Afghanistan ($N = 4$) and India ($N = 2$). Their educational backgrounds varied, ranging from primary education to higher education.

The focus group transcripts were analysed using thematic analysis, which facilitated the identification of recurring patterns within the data. Both inductive and deductive coding methods were employed to ensure a comprehensive analysis of the text. To maintain consistency, each member of the research team reviewed the transcripts independently and worked collaboratively to establish the final analytical framework.

Figure 1 presents a word cloud generated from the focus group discussion, highlighting key themes and concepts raised by participants. Prominent terms such as 'asylum', 'family', 'police', 'stay', 'safety', and 'education' reflect the central concerns shaping refugees' experiences and perceptions. The visibility of words like 'Taliban', 'fight', and 'beat' underscores the violence and persecution many participants fled, while references to 'school', 'work', and 'good' suggest aspirations for integration and stability.

The thematic analysis revealed three overarching themes that offer insights into different stages of the migration process. The first theme, 'Motivations and Drivers for Migration', encompasses the pre-migration phase, exploring the reasons behind the decision to migrate and preparations undertaken by individuals. The second theme, 'Experiences and Challenges During the Journey to Europe', relates to the movement phase, highlighting the obstacles and hardships refugees faced during migration. The third theme, 'Reconsideration of Transit Country as a New Destination', focuses on the arrival and integration phase, examining the conditions refugees encountered in Romania and their future plans for settlement or onwards migration.

These themes were developed using a deductive approach that was guided by the push and pull model (Lee 1966) providing a theoretical framework for systematically examining the data, along with the findings on changing preferences from our regression analysis. Push factors included, for example, conflict or persecution and pull factors included safety or economic opportunities influencing refugees' decisions. Alongside this, an inductive approach was employed to uncover new insights that emerged organically from the data. This dual approach enabled the development of a series of categories and codes that contributed to a more nuanced understanding of the migration process.

Figure 2 presents a hierarchical framework that visually organises the themes, sub-themes, categories, and codes derived from the data.

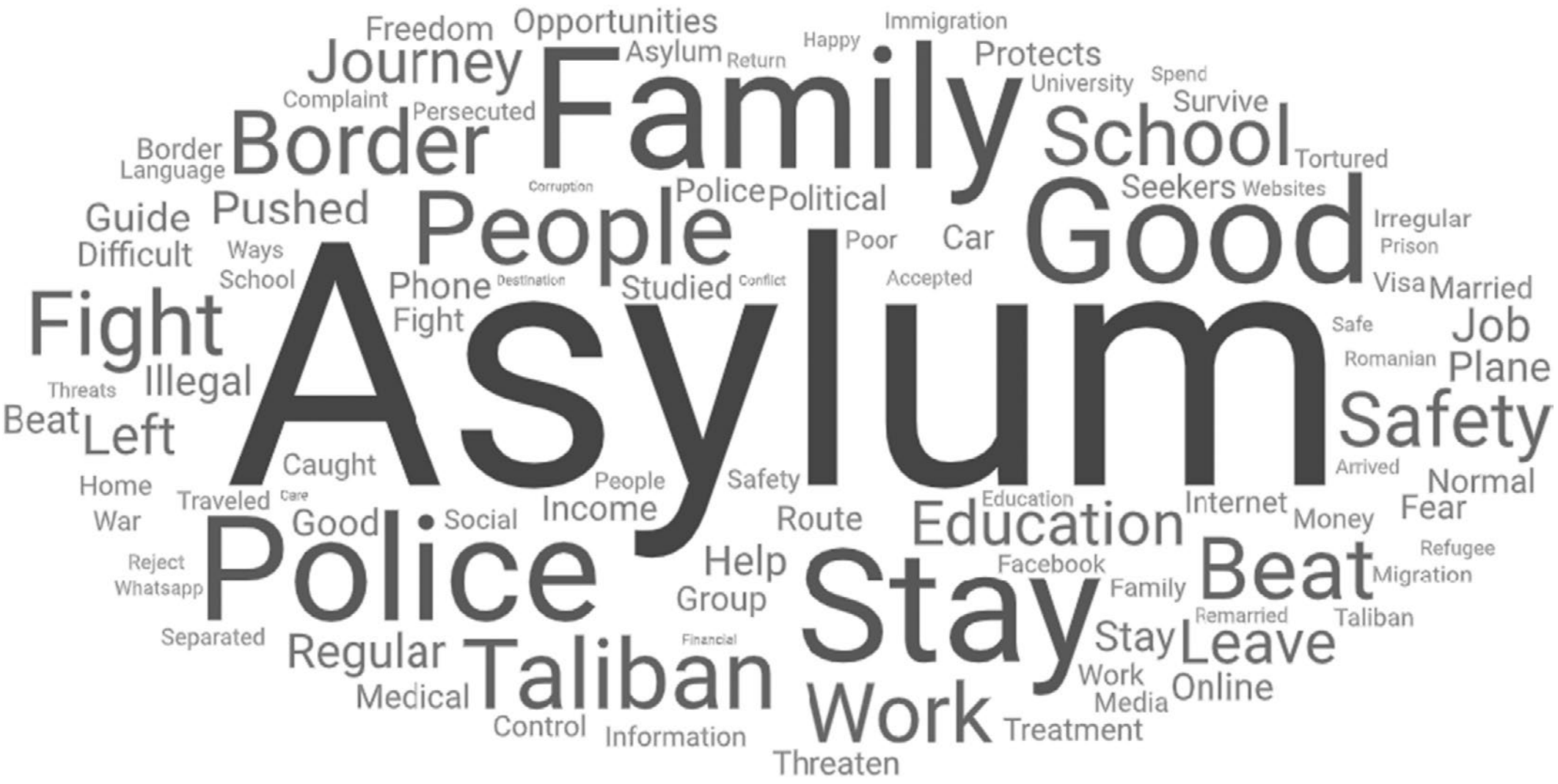

**FIGURE 1** | The most used words in focus groups. *Source:* Own elaboration based on focus group data.

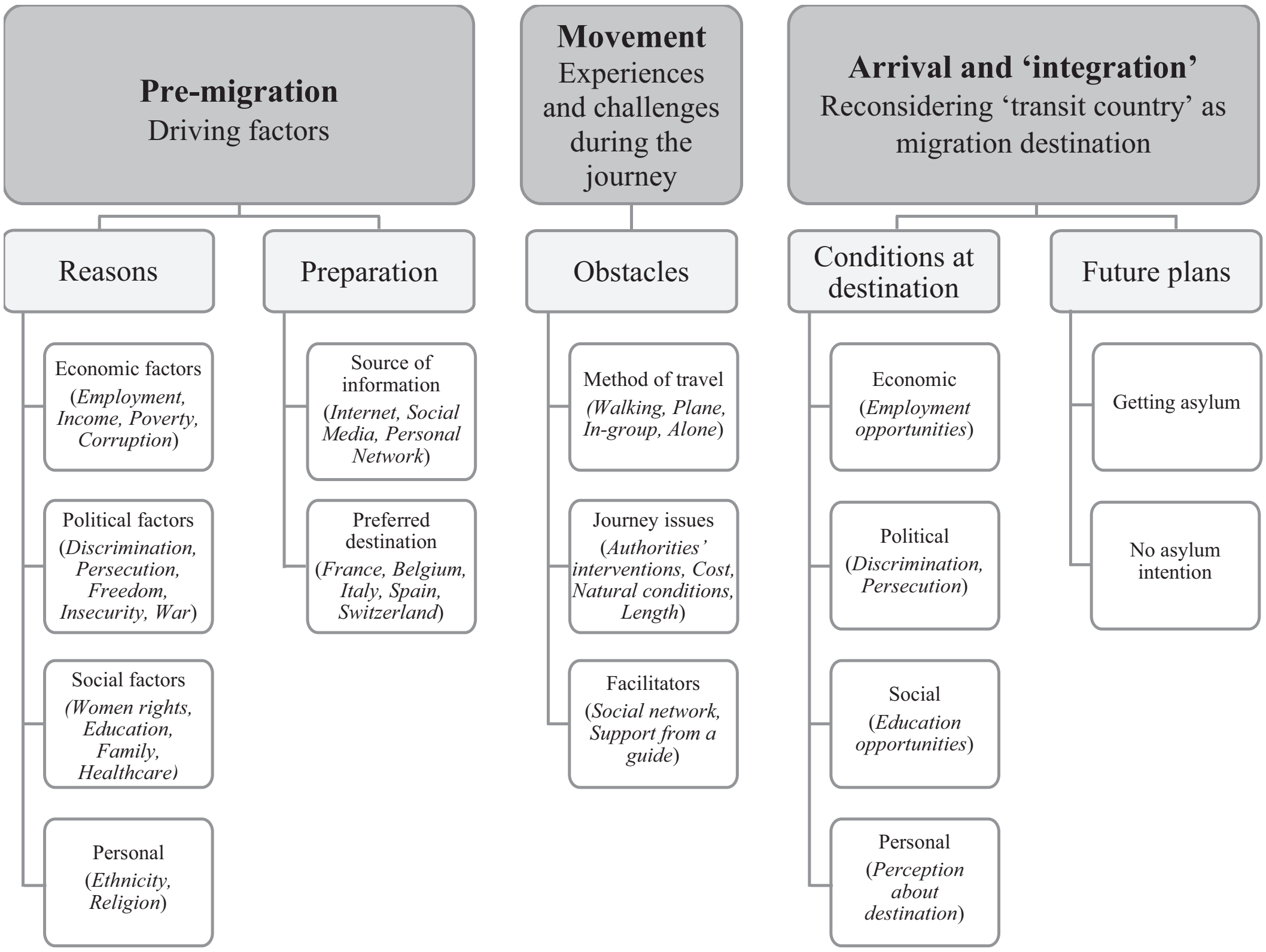

**FIGURE 2** | Themes, categories and codes emerged from data. *Source:* Own elaboration based on focus group data.

The analysis aims to establish causal relationships by exploring the motivations, decisions, and adaptations refugees make during their migration paths. As no substantial differences were observed in the topics discussed between the two focus groups underlying the qualitative analysis, the results presented in the following sections combine the findings from both groups. To manage readability, direct quotes are provided in Appendix S3 rather than in-line, with each quote referenced by its Appendix identifier (e.g., Appendix S3, Quote 1).

#### 4.2.1 | Pre-Migration Phase—Motivations and Drivers for Migration

In the pre-migration phase, focus group discussions revealed a complex interplay of political, social, and economic push factors motivating refugees to leave their countries of origin. Participants from Afghanistan and Pakistan described severe political instability and threats to personal safety as primary drivers. Many highlighted the pervasive influence of insurgent groups (e.g., the Taliban) that exert violent control over local populations, including the forced recruitment of young men and lethal punishment for non-compliance (Appendix S3, Quote 1). One Afghan participant detailed the compounded hardships under Taliban rule, from the lack of basic services like education and healthcare to the oppressive restrictions on religious freedom and women's mobility (Appendix S3, Quote 2). These testimonies underscore how conflict, persecution, and the collapse of normal life courses (employment, education, security) become intolerable, leaving people little choice but to flee. This supports the broader migration literature and our participants' backgrounds: many came from areas with high unemployment and low living standards, which likely contributed to their decision to leave. In essence, for individuals fleeing conflict zones, the imperative of survival and safety often outweighs economic considerations. In such contexts, the primary goal becomes securing a place of refuge, even if it means setting aside other preferences or long-term aspirations.

By contrast, participants from India, coming from a relatively more stable political context, emphasised economic challenges as their main driver for migration. One Indian participant, for example, explained that he had pursued education abroad and returned home only to find limited job prospects; he noted that if sufficient financial opportunities had existed in India, he would not have left (Appendix S3, Quote 3). For him, issues of personal safety or political freedom were not pressing, illustrating how human capital considerations (such as seeking better employment matching one's skills) can dominate in the absence of acute conflict. This distinction suggests that refugees' initial migration motives may differ based on origin: those from war-torn regions are pushed out by violence and persecution, whereas those from more peaceful but economically stagnant regions are primarily pulled by the promise of better livelihoods abroad. Despite this variation in underlying motives, however, destination preferences among participants converged in striking ways.

Nearly all respondents, regardless of country of origin, initially aspired to reach Western European countries, frequently mentioning Italy, France, and especially Germany as preferred destinations. These countries were perceived to offer better economic opportunities, higher living standards, and safety. Notably, at this stage, Romania was not widely perceived as an intentional or final destination; rather, it was predominantly viewed as a transit point along the way to Western Europe. Participants' initial knowledge of prospective destinations was significantly influenced by their social networks and exposure to digital media platforms. Many relied on information from friends or family already living in Europe and on social media groups to guide their plans (Appendix S3, Quote 4). Such network-informed decision making likely reinforced popular narratives of Western Europe as the ultimate goal, while countries like Romania were scarcely considered due to limited information and lack of existing diaspora connections.

The findings demonstrate that the pre-migration phase is characterised by refugees formulating migration plans influenced by push factors, such as insecurity or poverty in their home countries, and pull factors, notably the perceived prosperity and safety of prominent destination countries. Furthermore, refugees' expectations are mediated by their social networks, which disseminate both accurate information and misconceptions about potential destinations. However, as subsequent sections illustrate, these initial plans were substantially revised once migrants embarked on their journeys and confronted emerging realities.

#### 4.2.2 | Movement Phase—Experiences and Challenges During the Journey to Europe

The movement phase of the refugee journey involved passage through multiple countries and was fraught with well-documented challenges. All participants followed variations of the so-called 'regular' overland migration route toward Europe, typically trekking on foot across Iran, Turkey, and the Balkans (often via Greece or Bulgaria, then North Macedonia or Serbia, before reaching Romania). An Afghan refugee recounted that their group traversed at least six countries on foot on their way to Romania (Appendix S3, Quote 5). Such journeys were not only lengthy (often taking several months) but also extremely costly and hazardous. On average, participants reported spending around \$5000–\$6000 (USD) on the journey, (in survey data roughly 37% of respondents reported costs over \$5000). These expenditures, often financed through life savings or debt, underscore how refugees invest substantial human and financial capital into reaching Europe, an investment they later feel compelled to justify by eventually attaining stability and opportunity. The routes themselves presented severe physical challenges: refugees faced harsh terrain and weather that, in some instances, turned deadly. One participant described how two people in his travelling group perished from exposure to freezing temperatures while crossing snow-covered mountains early in the journey (Appendix S3, Quote 6). These qualitative accounts underscore the cumulative risk concept raised by Becker and Ferrara (2019), giving substance to the 'migration risks' that our quantitative model attempted to proxy through variables like conflict at origin or number of transit countries. The trauma and cost described here help explain why some refugees might reconsider their plans after such an ordeal.

Encounters with human dangers were also prevalent. Several participants reported instances of violence and abuse, particularly at border crossings. For example, one young man from Pakistan recounted being physically assaulted by border police when attempting to enter Romania from Serbia, describing beatings by both Serbian and Romanian officers (Appendix S3, Quote 7). Such experiences of brutality at the hands of authorities are unfortunately consistent with other evidence on the Western Balkan route. However, it is important to acknowledge that asylum seekers might minimise or understate these negative experiences when recounting them. Research on vulnerable populations (Haynes et al. 2023) suggests that victims of violence or abuse often downplay trauma as a coping mechanism or out of fear of jeopardising their asylum claims. In our focus groups, participants tended to frame incidents of police mistreatment as isolated 'unfortunate' events rather than evidence of systematic abuse, a narrative strategy noted in prior studies of refugee testimonies (Stewart 2005). This tendency to minimise hardship could stem from fear of retaliation, distrust in whether speaking out will lead to any change, or simply a psychological strategy to process trauma. Thus, while the focus group narratives confirm that many refugees faced violence and degrading treatment during transit, the true extent of these abuses may be under-reported in their accounts.

Despite the difficulties encountered during their journey, from hazardous landscapes to human violence, participants generally viewed the gruelling transit as a necessary means to an end. Several expressed a sense of relief upon finally reaching Romania, which for many was the first point in Europe where they felt relatively safe from immediate danger. One participant noted that after crossing into Romanian territory, authorities placed them in quarantine (due to COVID-19 measures) and provided food and shelter, and that 'all was OK' at that point (Appendix S3, Quote 14). This sentiment of relief suggests that, by the time they arrived in Romania, refugees were acutely aware that surviving the journey itself was a major accomplishment. Indeed, enduring the transit ordeal may alter refugees' calculus of where to settle after such hardship, the prospect of stopping in a country that offers basic safety and protection can become far more appealing than it seemed before departure. In other words, risk exposure during transit can trigger a reassessment of one's destination priorities, a clear manifestation of bounded rationality in migration decision-making. Refugees continuously update their plans based on the trade-offs between the dangers of moving onward and the opportunities available where they are.

Another notable finding in the movement phase is the variability of migration strategies employed by different refugees, illustrating their adaptability. While most travelled overland on foot, some with sufficient resources took alternative paths, such as flying part of the way. For instance, one participant from India managed to fly to Russia on a tourist visa as the first leg of his journey before continuing overland toward Europe (Appendix S3, Quote 8). Another described an entirely overland journey through Iran, Turkey, Bulgaria, Serbia and then into Romania (Appendix S3, Quote 9). These examples show that refugees are not homogeneously following a single script; rather, they make pragmatic choices based on the options available to them.

The quantitative analysis, which primarily enumerated the countries traversed, did not sufficiently capture these strategic complexities. The qualitative findings thus contribute significantly to our understanding by emphasising the critical roles of individual initiative and adaptability even within constrained contexts. Migrants strategically mobilised available resources, whether financial capital, visa opportunities, or social networks, to negotiate and facilitate their journeys. Such adaptive behaviours observed during transit have implications for refugees' subsequent settlement decisions. By the conclusion of the migration phase, participants had already incurred substantial costs and risks to arrive in Romania, a country initially perceived merely as a transit point. Upon arrival, however, migrants confronted the pivotal decision of remaining in this transit country or continuing their journey onward. Notably, experiences accumulated during transit reshaped participants' perspectives, making the prospect of settling temporarily in a location offering relative security more viable than initially anticipated.

#### 4.2.3 | Arrival and "Integration Phase"—Reconsideration of Transit Country as New Destination

The arrival and initial integration phase focuses on refugees' reassessment of Romania as a potential long-term destination. Upon arrival, a striking finding was that many participants expressed a genuine willingness to stay in Romania if they could secure asylum (legal protection) there. Initially, most had limited knowledge about Romania and did not plan to remain in it. Yet, having arrived, they recognised the value of stability and safety that Romania could offer. Legal status emerged as a crucial factor: one participant stated that he wished to stay in Romania and build a life there, but everything hinged on whether he would be granted protection; if his asylum claim was rejected, he would consider moving on to Italy (Appendix S3, Quote 10; Appendix S3, Quote 11). These reflections underscore that access to asylum and the guarantee of not being returned to danger are fundamental preconditions for refugees to invest in a country. This finding aligns closely with our quantitative results, where prioritising safety (over purely economic factors) was a key predictor of refugees switching their preference to the transit-country host.

Beyond legal protection, the social and economic environment in Romania played a pivotal role in whether participants viewed it as an acceptable destination. Several refugees voiced optimism about opportunities to work and integrate into the local society. One participant, for instance, was unsure about his right to work as an asylum seeker but was eager to find employment if possible, and he emphasised the welcoming interactions he had experienced with ordinary Romanian people on the street (Appendix S3, Quote 12). He noted that locals had been kind and never made derogatory remarks about his nationality or background. This and similar testimonials about the hospitality of the host society indicate that positive daily encounters can significantly influence refugees' perceptions of a country. Indeed, the absence of discrimination and the presence of local kindness were consistently noted by participants as a refreshing change compared to experiences in other countries along the journey (Appendix S3, Quote 13). This positive initial reception indicates that social factors, not just pre-existing ties but new connections and a welcoming community, can influence refugees' willingness to remain. In effect, some participants were beginning to build the very networks that might anchor them in Romania, complementing the 'safety' motive quantified in our survey. Such social acceptance resonates with the quantitative finding that favourable asylum conditions (which encompass not just official policies but also the general social climate) are associated with a higher likelihood of refugees considering a transit country for settlement.

It is important to contextualise these positive experiences within Romania's broader position as an emerging country of immigration. While individual refugees in our study expressed appreciation for the friendliness of Romanian people, existing literature emphasises that Romania has not historically served as a primary destination for refugees, and its integration infrastructure remains in a developmental phase. Traditionally characterised as a country of emigration rather than immigration, Romania's policies aimed at immigrant integration, such as language acquisition programmes and employment support services, are still relatively underdeveloped (Bejan 2021). For those who choose to remain, perceived safety and societal hospitality frequently emerge as key motivators, a sentiment echoed by participants in this study. However, structural barriers, including limited interaction with broader Romanian society beyond NGO-operated facilities and challenges related to the recognition of foreign professional qualifications continue to constrain opportunities for full integration. Focus group participants, most of whom had spent the majority of their brief stay in Romania within reception centres or under the care of non-governmental organisations, reported interacting with locals primarily through informal or incidental encounters. As such, their overwhelmingly positive characterisations of Romanians as 'good and not mean' to foreigners (Appendix S3, Quote 13) are likely reflective of initial, and often idealised, impressions. Whether this favourable perception endures over the course of long-term integration remains an open question beyond the scope of this study. Nonetheless, in the early stages of settlement, the welcoming social climate in Romania appeared to enhance the perceived viability and appeal of remaining in the country.

Crucially, participants' individual aspirations become particularly salient during the arrival phase. Far from seeing themselves as passive victims or transient outsiders, many refugees demonstrated a strong desire to contribute to Romanian society and rebuild their lives there if given the chance. They spoke about plans and hopes for the future, finding jobs, pursuing education, and reuniting with family which highlight a proactive attitude toward integration. For instance, multiple individuals expressed determination to work and even to continue their studies or vocational training in Romania, despite the interruptions in their education due to displacement (Appendix S3, Quotes 15–17). One participant talked about wanting to attend school and gain an education even at an older age, another about aspirations to become a chef, and another about bringing his family to Romania when possible. These forward-looking ambitions reflect a desire among refugees to invest in their own development through education, training, or skill utilisation in ways that can contribute to the host country's economy, provided that enabling opportunities are made available. The articulation of such goals by participants suggests that, once immediate safety had been secured, their focus shifted toward long-term integration prospects and active participation in the host society. In other words, their priorities evolved from escaping danger to pursuing stability and personal development. This evolution underscores refugees' adaptability and resilience. It also complements the quantitative evidence by adding a human dimension: while the survey data identified 'asylum conditions' and 'safety' as significant factors for choosing to stay in Romania, the qualitative narratives show *how* those factors translate into real intentions and efforts on the ground. Refugees in Romania are not simply pausing their journeys; rather, many are actively seeking to transform a temporary transit stop into a place of settlement by entering the labor market, acquiring new skills or language proficiency, and establishing a sense of permanence within the host society.

Taken together, the qualitative findings offer a nuanced understanding of how Romania, although not initially perceived as a primary destination by the refugees, emerged as a contingent site of settlement under favourable conditions. The focus group narratives demonstrate that when a transit country affords a sufficient sense of physical and legal security, coupled with social acceptance and opportunities for socio-economic participation, refugees may reassess their trajectories and consider permanent settlement.

## 5 | Conclusions

This study provides an in-depth analysis of the factors influencing how refugees' destination preferences evolve during transit. By employing a mixed-methods design that integrates quantitative data from the IOM's Flow Monitoring Survey and qualitative insights from focus groups conducted in Romania, we explore both the structural and personal factors shaping refugee decision-making. Our findings emphasise that migration decisions are dynamic and contingent, influenced by macro-level drivers such as conflict and political instability, as well as micro-level factors like individual safety concerns, social networks, and perceived opportunities in transit countries.

The quantitative analysis revealed that safety and legal stability are among the most significant predictors of preference change, particularly in the Romanian context. While initial intentions often focus on reaching Western Europe, many refugees reassess their plans during transit, especially when they encounter relatively favourable conditions in countries traditionally seen as waypoints. This supports theoretical models such as bounded rationality (Brunarska 2019; Simon 1957) and risk minimisation (Becker and Ferrara 2019), which explain how refugees make 'good enough' decisions under uncertainty. Human capital also plays a role, more educated individuals in the pooled sample were more likely to change destination plans, although this trend was not evident in the Romania-specific analysis, suggesting that the salience of education may vary with local conditions and opportunity structures.

The qualitative findings add depth to these patterns, capturing the lived realities behind migration decisions. Participants described how hardships during transit, including violence, financial strain, and emotional stress, reshaped their perceptions of viable destinations. Upon arrival, legal protection, safety, and positive social encounters emerged as critical factors in transforming perceptions of Romania from a mere transit country into a potential destination. Refugees also expressed a strong desire for integration and self-reliance, often aspiring to work, study, and reunite with family.

Taken together, these findings underscore the importance of viewing destination selection as a fluid, context-sensitive process rather than a fixed goal. Refugees continuously revise their plans in response to both structural constraints and new opportunities encountered en route. The study's insights have important policy implications. Transit countries can become meaningful destinations when they offer not only safety and access to asylum but also clear pathways to integration. Accelerating asylum procedures, ensuring fairness, and providing legal and social protections may encourage asylum seekers to remain and integrate, thereby easing pressure on traditional destination countries. However, retaining highly educated or skilled individuals will require more targeted efforts, including investments in language training, recognition of foreign qualifications, and access to suitable employment. At the same time, policymakers must acknowledge that societal attitudes also play a key role. Cultural proximity, such as that perceived between Ukrainian refugees and host communities, can influence public acceptance, but inclusive policies, civic engagement, and equitable treatment of all refugee groups are crucial for building sustainable integration pathways. As Jacobsen and Landau (2003) highlight, host perceptions are shaped by a complex interplay of economic concerns, cultural narratives, and political positioning, all of which can either facilitate or hinder refugee integration.

Nonetheless, this study acknowledges several important limitations. Refugees' precarious legal status may have influenced their responses, particularly in the qualitative phase, where participants may have minimised negative experiences due to their vulnerable position and the inherent power dynamics with authorities (Haynes et al. 2023; Stewart 2005). Furthermore, the absence of female participants represents a notable limitation, as migration experiences are highly gendered. Female asylum seekers often face distinct barriers and vulnerabilities, especially in relation to safety, legal recognition, and integration into host societies (Van Wetten et al. 2001). Future studies should include more gender-balanced samples and explore how these dynamics affect migration trajectories and outcomes across different groups.

Finally, the study contributes to theoretical debates by challenging the assumption of fixed destination preferences. By applying bounded rationality, human capital theory, and network theory, it presents a comprehensive framework for understanding how refugees adapt to shifting realities during migration. Empirically, it offers rare insights into the under-researched transit countries of Central and Eastern Europe, particularly Romania, highlighting their evolving role in Europe's asylum landscape.

In sum, this research challenges the notion that 'transit' countries are merely waystations. Under the right conditions, they can become meaningful destinations in their own right.

## Acknowledgements

Open access publishing facilitated by Anelis Plus (the official name of "Asociatia Universitatilor, a Institutelor de Cercetare - Dezvoltare si a Bibliotecilor Centrale Universitare din Romania"), as part of the Wiley - Anelis Plus agreement.

## Conflicts of Interest

The authors declare no conflicts of interest.

## Data Availability Statement

Data sharing is not applicable to this article as no new data were created or analyzed in this study.

## Peer Review

The peer review history for this article is available at https://www.webofscience.com/api/gateway/wos/peer-review/10.1111/imig.70066.

## Endnotes

[1] For example, the education level can influence the ability of refugees to integrate in the labour market at destination.

[2] Immigrants have the right to apply for asylum either when they already are on the territory of Romania or at the border. When a person expresses the intention to seek asylum (s)he is sent to the Regional Centres for Asylum Seekers. In a maximum 9 months they get a resolution which (a) grants a form of protection; (b) grants access to the territory if the application is not manifestly unfounded or if there are indications that Dublin or Admissibility grounds apply; or (c) rejects the application as manifestly unfounded and grants no access to the territory.

[3] LOGSS ('Grup de Inițiative Sociale') is an accredited local NGO that provides emergency material aid and comprehensive support services, including information, legal guidance, Romanian language classes, medical care, and psychological counselling to refugees, asylum seekers, and third-country nationals in Timișoara, fostering their integration in Romania.

### Supporting Information

Additional supporting information can be found online in the Supporting Information section.